\documentclass[aps,pra,twocolumn,amsmath,amssymb,nofootinbib,showpacs,superscriptaddress]{revtex4-1}
\usepackage[english]{babel}
\usepackage{latexsym}
\usepackage{graphics}
\usepackage{graphicx}
\usepackage{epsfig}
\usepackage{color}
\usepackage{bm}
\usepackage{amsmath}
\usepackage{amssymb}
\usepackage{amsthm}
\usepackage{dcolumn}
\usepackage{float}
\usepackage{hyperref}
\usepackage{color}
\usepackage{epstopdf}
\usepackage{cleveref}
\usepackage[svgnames]{xcolor}
\hypersetup{hidelinks,colorlinks=true,allcolors=DarkBlue}

\newcommand{\ket}[1]{|#1\rangle}
\newcommand{\bra}[1]{\langle#1|}

\begin{document}

\preprint{APS/123-QED}

\title{Information processing using three-qubit and \\ qubit--qutrit encodings of noncomposite quantum systems}
\author{A.A. Popov}
\affiliation{Bauman Moscow State Technical University, Moscow 105005, Russia}
\affiliation{Acronis Ltd, Moscow 127566, Russia}
\author{E.O. Kiktenko}
\affiliation{Theoretical Department, DEPHAN, Skolkovo, Moscow 143025, Russia}
\affiliation{Bauman Moscow State Technical University, Moscow 105005, Russia}
\author{A.K. Fedorov}
\affiliation{Acronis Ltd, Moscow 127566, Russia}
\affiliation{Russian Quantum Center, Skolkovo, Moscow 143025, Russia}
\affiliation{LPTMS, CNRS, Univ. Paris-Sud, Universit\'e Paris-Saclay, Orsay 91405, France}
\author{V.I. Man'ko}
\affiliation{P.N. Lebedev Physical Institute, Russian Academy of Sciences, Moscow 119991, Russia}
\affiliation{Moscow Institute of Physics and Technology (State University), Moscow Region 141700, Russia}

\date{\today}

\begin{abstract}
We study quantum information properties of a seven-level system realized by a particle in an one-dimensional square-well trap.
Features of encodings of seven-level systems in a form of three-qubit or qubit--qutrit systems are discussed.
We use the three-qubit encoding of the system in order to investigate subadditivity and strong subadditivity conditions for the thermal state of the particle.
The qubit--qutrit encoding is employed to suggest a single qudit algorithm for calculation of parity of a bit string.
Obtained results indicate on the potential resource of multilevel systems for realization of quantum information processing. 

\begin{description}
\item[PACS numbers]
03.65.Wj, 03.65.Ta, 42.50.Xa
\end{description}
\end{abstract}

\maketitle

\section{Introduction}

Design of universal large-scale quantum computers is one of the most challenging problems of quantum information technologies~\cite{Nielsen}.
Quantum computing devices would allow one to solve certain mathematical problems in a more efficient way~\cite{Nielsen,Feynman,Ladd} in compare with their classical counterparts,
in particular, tasks dealing with integer factorization and discrete logarithm problems~\cite{Shor}, or searching an unsorted database~\cite{Grover}.

A common way for building of quantum computing devices is to create a scalable system of qubits, which play a role of information units~\cite{Nielsen}.
Qubits can be realized using a variety of different physical systems including photons, nitrogen-vacancy centers, superconducting circuits, ultracold trapped atoms, ions, and molecules.

Recently, 
multilevel ({i.e.}, noncomposite) quantum systems attracted a significant deal of interest as a potential platform for quantum 
technologies~\cite{MAManko1,MAManko2,MAManko3,MAManko4,MAManko5,MAManko6,MAManko7,MAManko8,
MAManko9,MAManko10,MAManko11,MAManko12,MAManko13,MAManko14,MAManko15,Kiktenko,Kiktenko2,Kiktenko3,Kiktenko4,Gedik}.
On the one hand, such multilevel quantum systems can be realized using a single physical system, {e.g.}, multilevel superconducting artificial atoms~\cite{Katz,Katz2,Gustavsson,Katz3,Ustinov}. 
On the other hand, noncomposite systems `contain' a number of virtual subsystems.
It has been shown that hidden quantum correlation between virtual subsystem can be quantified~\cite{MAManko5}. 

Speaking in terms of the quantum Shannon theory~\cite{Nielsen}, 
one can establish a direct correspondence between information and entropic measures for composite and noncomposite quantum systems.
Comprehensive studies of information and entropic characteristics of noncomposite quantum system demonstrated their resource for information 
processing~\cite{MAManko1,MAManko2,MAManko3,MAManko4,MAManko5,MAManko6,MAManko7,MAManko8,
MAManko9,MAManko10,MAManko11,MAManko12,MAManko13,MAManko14,MAManko15,Kiktenko,Kiktenko2,Kiktenko3,Kiktenko4}. 
In particular, the single qudit version of the Deutsch algorithm using through anharmonic superconducting multilevel artificial atom has been suggested~\cite{MAManko5}. 
It should be mentioned that this paradigm has much in common with the foundation of the Kochen--Specker theorem \cite{KochenSpecker} 
and pioneering experimental results demonstrating fundamentally non-classical properties of multilevel systems~\cite{Zeilinger}.

\begin{figure}
	\begin{center}
		\includegraphics[width=0.475\linewidth]{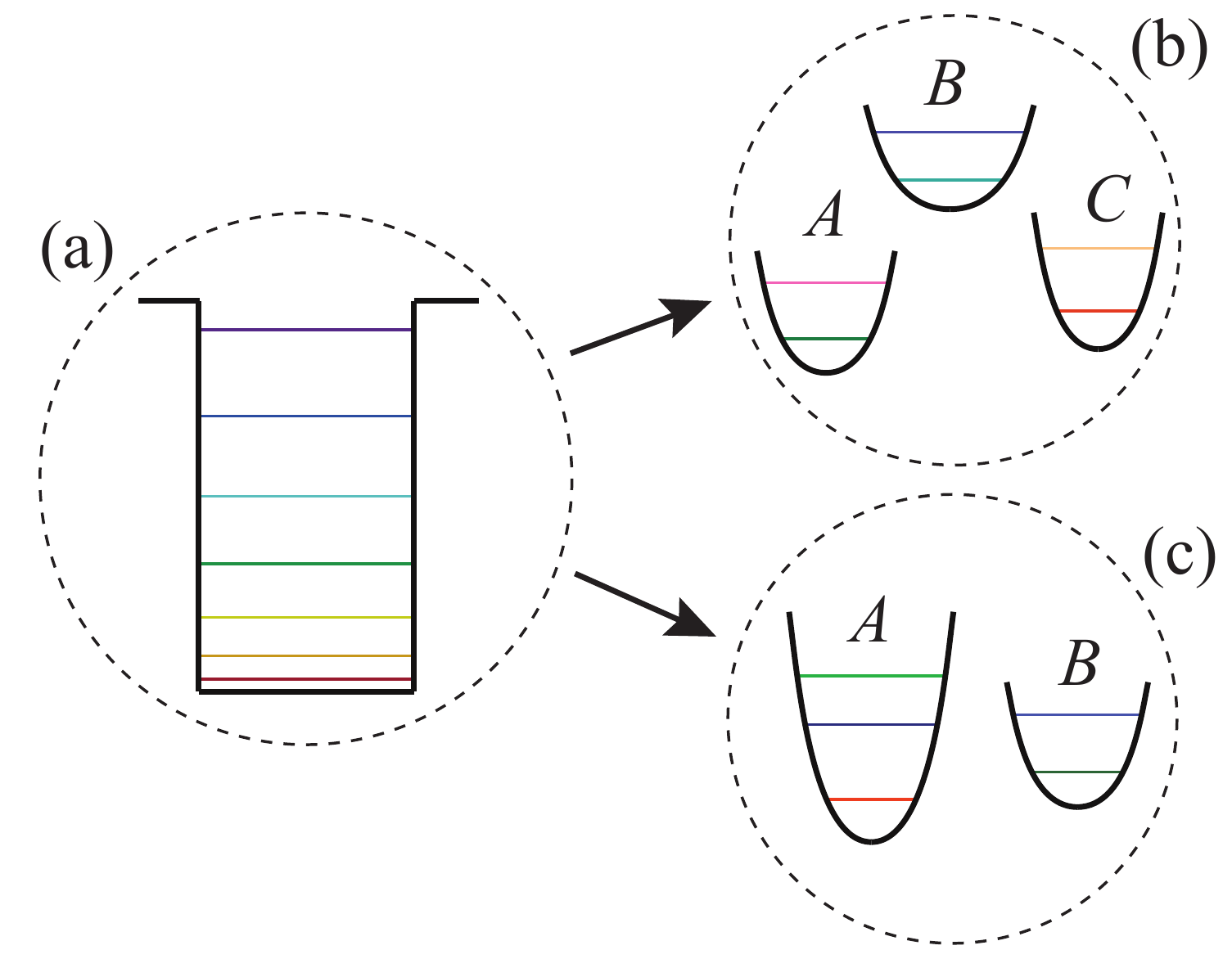}
	\end{center}
	\vskip -8mm
	\caption{Representation of the space of a particle in a 1D square-well trap with seven bound states (a) as there-qubit space (b) and qubit--qutrit space (c).}
	\label{fig:potwell}
\end{figure}

In the present work, 
we consider a seven-level quantum system as a platform for realization of quantum information processing.
Interesting peculiarity of this system is possibility to represent it using two significantly different types of composite representation ({i.e.}, encoding):
three-qubit or qubit--qutrit systems (see Fig.~\ref{fig:potwell}).
We employ the three-qubit encoding in order to study subadditivity and strong subadditivity conditions.
The qubit--qutrit encoding is used in order to suggest a single qudit algorithm for bit-string parity-calculation.
As a possible experimental setup, we consider a particle in a one-dimensional (1D) square-well trap,
which can be realized using currently available experimental tools. 

Our paper is organized as follows.
In Sec.~\ref{sec:setup}, we describe a general setup for study seven-level quantum systems using a particle in 1D square-well trap.
In Sec.~\ref{sec:three-qubit}, we use the three-qubit encoding of seven-level quantum systems for investigate subadditivity and strong subadditivity conditions. 
We suggest a single qudit algorithm for parity-calculation using the qubit--qutrit encoding in Sec.~\ref{sec:qubit--qutrit}.
We conclude our results in Sec.~\ref{sec:conclusion}.

\section{Setup}\label{sec:setup}

We recall the basic relations describing a particle in a 1D square-well trap. 
The Hamiltonian of the system is as follows~\cite{Griffiths}:
\begin{equation}\label{eq:hamiltonian}
H = -\frac{\hbar^2}{2m} \frac{\partial^2}{\partial x^2} + V(x),
\end{equation}
where the potential is not equal to zero in the finite domain:
\begin{equation}
V(x)=\left\{\begin{array}{cc}
0& |x| > a/2\\
-V_{0} & |x| \leq a/2
\end{array}\right.,
\end{equation}
$m$ is a mass of particle, $x$ is its coordinate, and $a$ is a width of the well. 
The number of bounded states $N$ is determined by the condition
\begin{equation}
\frac{\pi}{a}(N-1)\le{K}\le\frac{\pi}{a}N, \quad K=\frac{\sqrt{2mV_0}}{\hbar}.
\end{equation}
In what is presented below we use dimensionless units $\hbar=k_{\rm B}=a=m=q=1$, where $k_{\rm B}$ is the Boltzmann constant and $q$ is a charge of the particle.
In order to realize the system with seven bounded states, we consider a specific value for the potential depth $V_0=-200$.

In general case, the density matrix of the seven-level system has the following form:
\begin{equation}
\rho=\begin{bmatrix}
\rho_{{1,1}}&\rho_{{1,2}}&\rho_{{1,3}}&\rho_{{1,4}}&\rho_{{1,5}}&\rho_{{1,6}}&\rho_{{1,7}}\\
\noalign{\medskip}\rho^*_{{1,2}}&\rho_{{2,2}}&\rho_{{2,3}}&\rho_{{2,4}}&\rho_{{2,5}}&\rho_{{2,6}}&\rho_{{2,7}}\\
\noalign{\medskip}\rho^*_{{1,3}}&\rho^*_{{2,3}}&\rho_{{3,3}}&\rho_{{3,4}}&\rho_{{3,5}}&\rho_{{3,6}}&\rho_{{3,7}}\\
\noalign{\medskip}\rho^*_{{1,4}}&\rho^*_{{2,4}}&\rho^*_{{3,4}}&\rho_{{4,4}}&\rho_{{4,5}}&\rho_{{4,6}}&\rho_{{4,7}}\\
\noalign{\medskip}\rho^*_{{1,5}}&\rho^*_{{2,5}}&\rho^*_{{3,5}}&\rho^*_{{4,5}}&\rho_{{5,5}}&\rho_{{5,6}}&\rho_{{5,7}}\\ 
\noalign{\medskip}\rho^*_{{1,6}}&\rho^*_{{2,6}}&\rho^*_{{3,6}}&\rho^*_{{4,6}}&\rho^*_{{5,6}}&\rho_{{6,6}}&\rho_{{6,7}}\\
\noalign{\medskip}\rho^*_{{1,7}}&\rho^*_{{2,7}}&\rho^*_{{3,7}}&\rho^*_{{4,7}}&\rho^*_{{5,7}}&\rho^*_{{6,7}}&\rho_{{7,7}}
\end{bmatrix}
\end{equation}

The seven-dimensional Hilbert space $\mathcal{H}$ of such system  can not be decomposed as a product of spaces of lower dimension. 
However, there are two possibilities: 
(i) to extend the space to equivalent eight-dimensional by introducing new fictitious level, 
and (ii) to operate the six-dimensional subspace using the seventh level as auxiliary degree of freedom.

\section{Three-qubit encoding: subadditivity}\label{sec:three-qubit}

We start our investigation of the system from representation of the Hilbert space of the entire system as the product of three abstract two-dimensional subspaces:
\begin{equation}
\mathcal{H}=\mathcal{H}_A\otimes\mathcal{H}_B\otimes\mathcal{H}_C.
\end{equation}
For this purpose, we consider the following encoding scheme:
\begin{equation}\label{eq:rerp1}
\begin{aligned}
&\ket{1}\leftrightarrow\ket{000}, \quad \ket{2}\leftrightarrow\ket{001}, \\
&\ket{3}\leftrightarrow\ket{010}, \quad \ket{4}\leftrightarrow\ket{011}, \\
&\ket{5}\leftrightarrow\ket{100}, \quad \ket{6}\leftrightarrow\ket{101}, \\
&\ket{7}\leftrightarrow\ket{110}, \quad \ket{\rm aux}\leftrightarrow\ket{111}.\\
\end{aligned}
\end{equation}
Here $\ket{n}$ denotes the $n$th energy eigenstate and $\ket{mkl}$ denotes the tensor product $\ket{m}_A\otimes\ket{k}_B\otimes\ket{l}_C$ of the logical qubits' states. 
The state $\ket{\rm aux}$ corresponds to fictitious unoccupied level used to extend the $7\times7$ density matrix of the system to $8\times8$ one.

The density matrices of the three-qubit, two-qubit and one qubit reduced states have the form:
\begin{equation}
\varrho_{ABC}=
\begin{bmatrix}
\rho_{{1,1}}&\rho_{{1,2}}&\rho_{{1,3}}&\rho_{{1,4}}&\rho_{{1,5}}&\rho_{{1,6}}&\rho_{{1,7}} & 0\\
\noalign{\medskip}\rho^*_{{1,2}}&\rho_{{2,2}}&\rho_{{2,3}}&\rho_{{2,4}}&\rho_{{2,5}}&\rho_{{2,6}}&\rho_{{2,7}}&0\\
\noalign{\medskip}\rho^*_{{1,3}}&\rho^*_{{2,3}}&\rho_{{3,3}}&\rho_{{3,4}}&\rho_{{3,5}}&\rho_{{3,6}}&\rho_{{3,7}}&0\\
\noalign{\medskip}\rho^*_{{1,4}}&\rho^*_{{2,4}}&\rho^*_{{3,4}}&\rho_{{4,4}}&\rho_{{4,5}}&\rho_{{4,6}}&\rho_{{4,7}}&0\\
\noalign{\medskip}\rho^*_{{1,5}}&\rho^*_{{2,5}}&\rho^*_{{3,5}}&\rho^*_{{4,5}}&\rho_{{5,5}}&\rho_{{5,6}}&\rho_{{5,7}}&0\\
\noalign{\medskip}\rho^*_{{1,6}}&\rho^*_{{2,6}}&\rho^*_{{3,6}}&\rho^*_{{4,6}}&\rho^*_{{5,6}}&\rho_{{6,6}}&\rho_{{6,7}}&0\\
\noalign{\medskip}\rho^*_{{1,7}}&\rho^*_{{2,7}}&\rho^*_{{3,7}}&\rho^*_{{4,7}}&\rho^*_{{5,7}}&\rho^*_{{6,7}}&\rho_{{7,7}}&0\\
0&0&0&0&0&0&0&0\\
\end{bmatrix},\\
\end{equation}
\begin{equation}
\varrho_{AB}=
\begin{bmatrix}
\rho_{1,1}+\rho_{2,2}&\rho_{1,3}+\rho_{2,4}&\rho_{1,5}+\rho_{2,6}&\rho_{1,7}\\
\noalign{\medskip}\rho_{1,3}^*+\rho_{2,4}^*&\rho_{3,3}+\rho_{4,4}&\rho_{3,5}+\rho_{4,6}&\rho_{3,7}\\ \noalign{\medskip}\rho_{1,5}^*+\rho_{2,6}^*&\rho_{3,5}^*+\rho_{4,6}^*&\rho_{5,5}+\rho_{6,6}&\rho_{5,7}\\
\noalign{\medskip}\rho_{1,7}^*&\rho_{3,7}^*&\rho_{5,7}^*&\rho_{7,7}
\end{bmatrix},
\end{equation}
\begin{equation}
\varrho_{AC}=
\begin{bmatrix}
\rho_{1,1}+\rho_{3,3}&\rho_{1,2}+\rho_{3,4}&\rho_{1,5}+\rho_{3,7}&\rho_{1,6}\\
\noalign{\medskip}\rho_{1,2}^*+\rho_{3,4}^*&\rho_{2,2}+\rho_{4,4}&\rho_{2,5}+\rho_{4,7}&\rho_{2,6}\\ \noalign{\medskip}\rho_{1,5}^*+\rho_{3,7}^*&\rho_{2,5}^*+\rho_{4,7}^*&\rho_{5,5}+\rho_{7,7}&\rho_{5,6}\\
\noalign{\medskip}\rho_{1,6}^*&\rho_{2,6}^*&\rho_{5,6}^*&\rho_{6,6}
\end{bmatrix}, \\
\end{equation}
\begin{equation}
\varrho_{BC}=
\begin{bmatrix}
\rho_{1,1}+\rho_{5,5}&\rho_{1,2}+\rho_{5,6}&\rho_{1,3}+\rho_{5,7}&\rho_{1,4}\\
\noalign{\medskip}\rho_{1,2}^*+\rho_{5,6}^*&\rho_{2,2}+\rho_{6,6}&\rho_{2,3}+\rho_{6,7}&\rho_{2,4}\\ \noalign{\medskip}\rho_{1,3}^*+\rho_{5,7}^*&\rho_{2,3}^*+\rho_{6,7}^*&\rho_{3,3}+\rho_{7,7}&\rho_{3,4}\\
\noalign{\medskip}\rho_{1,4}^*&\rho_{2,4}^*&\rho_{3,4}^*&\rho_{4,4}
\end{bmatrix}, \\
\end{equation}
\begin{equation}
\varrho_{A}{=}\begin{bmatrix}
\rho_{{1,1}}+\rho_{{2,2}}+\rho_{{3,3}}+\rho_{{4,4}}&\rho_{{1,5}}+\rho_{{2,6}}+\rho_{{3,7}}\\
\noalign{\medskip}\rho^*_{{1,5}}+\rho^*_{{2,6}}+\rho^*_{{3,7}}&\rho_{{5,5}}+\rho_{{6,6}}+\rho_{{7,7}}
\end{bmatrix},
\end{equation}
\begin{equation}
\varrho_{B}{=}
\begin{bmatrix}
\rho_{{1,1}}+\rho_{{5,5}}+\rho_{{2,2}}+\rho_{{6,6}} &\rho_{{1,3}}+\rho_{{5,7}}+\rho_{{2,4}}\\
\noalign{\medskip}\rho^*_{{1,3}}+\rho^*_{{5,7}}+\rho^*_{{2,4}}&\rho_{{3,3}}+\rho_{{7,7}}+\rho_{{4,4}}
\end{bmatrix},
\end{equation}
\begin{equation}
\varrho_{C}{=}\begin{bmatrix}
\rho_{{1,1}}+\rho_{{3,3}}+\rho_{{5,5}}+\rho_{{7,7}} &\rho_{{1,2}}+\rho_{{3,4}}+\rho_{{5,6}}\\
\noalign{\medskip}\rho^*_{{1,2}}+\rho^*_{{3,4}}+\rho^*_{{5,6}}&\rho_{{2,2}}+\rho_{{4,4}}+\rho_{{6,6}}
\end{bmatrix}.
\end{equation}
One can see that representation (\ref{eq:rerp1}) does not provide the full three-qubit basis. 
Therefore, it is impossible to carry out universal three-qubit computation in such a system. 

\begin{figure}
	\begin{center}
	\includegraphics[width=1\linewidth]{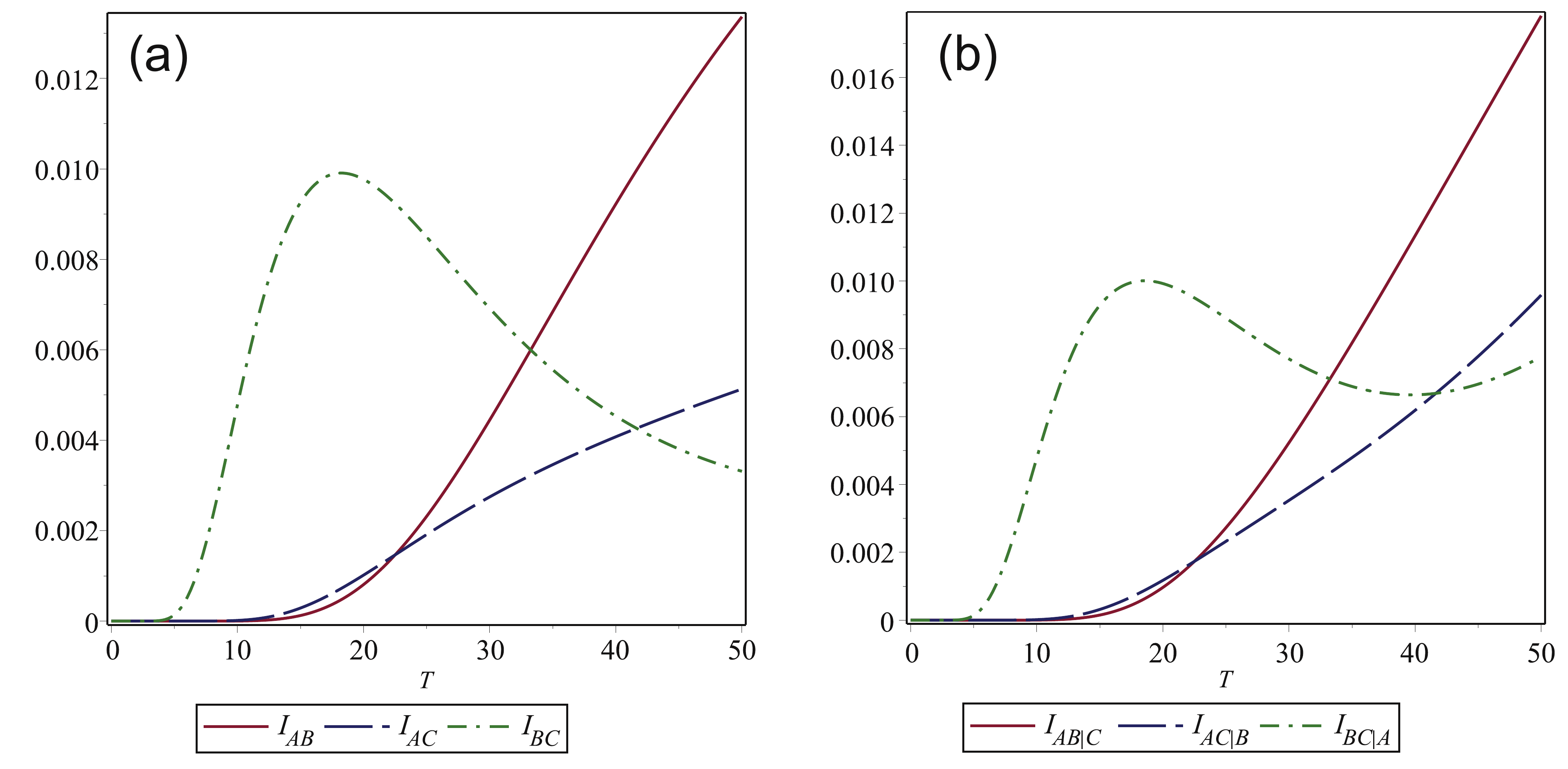}
	\end{center}
	\caption{Dependence of quantum mutual information (a) and quantum conditional mutual information (b) of subsystems on dimensionless temperature.}
	\label{fig:information}
\end{figure}

Considering the state of the non composite system as a state of three qubits we can introduce the 
quantum mutual information and conditional quantum mutual information as follows:
\begin{eqnarray}
&I_{XY}=S[\varrho_{X}]+S[\varrho_{Y}]-S[\varrho_{XY}], \label{eq:MInf} \\ 
&I_{XY|Z}=S[\varrho_{XZ}]+S[\varrho_{YZ}]-S[\varrho_{Z}]-S[\varrho_{XYZ}], \label{eq:CMInf}
\end{eqnarray}
where $X,Y,Z\in\{A,B,C\}$ and $S[\sigma]$ is von Neumann entropy:
\begin{equation}
S[\sigma]=-\textrm{Tr}[\sigma\;\textrm{log}_{2} \sigma ].
\end{equation}
The values of~\eqref{eq:MInf} and \eqref{eq:CMInf} always turn to be nonnegative that is known as subadditivity~\cite{SubAdd} and strong subadditivity conditions~\cite{StrongSubAdd}

We can check the positivity of mutual information and conditional mutual information constructed for virtual subsystems. 
As an example, we consider the thermal state, whose density matrix is described by the Gibbs distribution:
\begin{equation}
	\rho = \frac{1}{Z}\exp\left(-\frac{H}{T}\right), \quad Z={\rm Tr}\left[ \exp\left(-\frac{H}{T}\right) \right],
\end{equation}
where $T$ is the dimensionless temperature and $Z$ is the partition function. 

The results for all possible partitions are presented in Fig.~\ref{fig:information} (to exclude the contribution of continuous spectrum in density matrix, we consider only temperatures below the critical value).
One can see that the growth of the temperature from the zero value leads to appearance of correlations between virtual qubits.
At the same time, the mutual information of the partition $AC$ demonstrates nontrivial non monotonic behaviour having an extremum at $T\approx18$.

\section{qubit--qutrit encoding: Parity check algorithm}\label{sec:qubit--qutrit}

\begin{figure*}
	\begin{center}
			\includegraphics[width=1\linewidth]{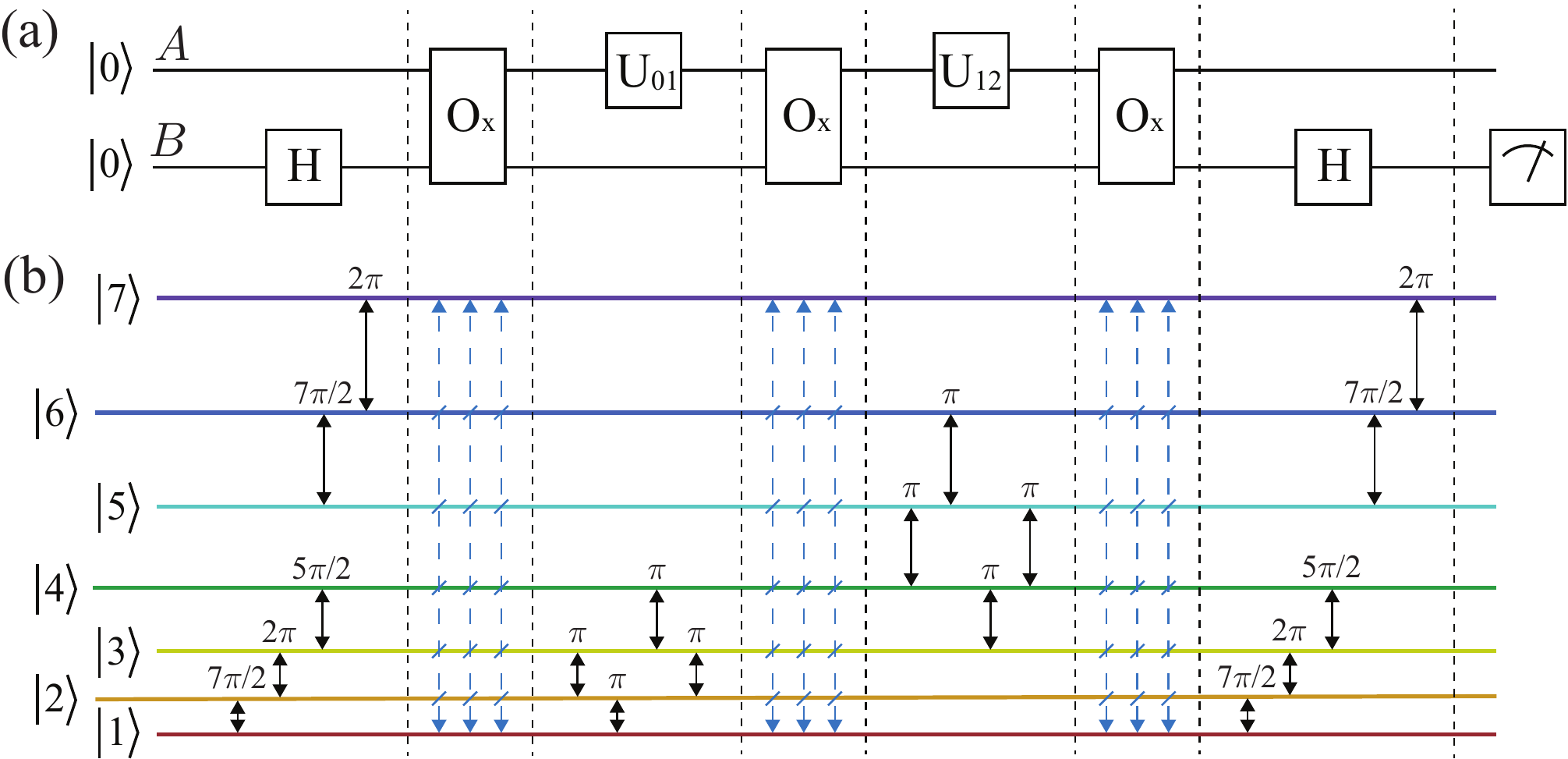}
	\end{center}
	\vskip -6mm
	\caption{The quantum circuit of the parity check algorithm (a) and corresponding sequence of pulses acting a particle in a square trap (b).}
	\label{fig:algorithm}
\end{figure*}

In order to get an insight about possibility to use this system for universal quantum computations, 
we can refer to the six-dimensional subspace, which can be represented as a product of spaces of one qutrit $A$ and one qubit $B$:
\begin{equation}
\mathcal{H}=\mathcal{H}_A\otimes\mathcal{H}_B.
\end{equation}
This representation is defined by the following encoding: 
\begin{equation}\label{eq:repr2}
\begin{aligned}
&\ket{1}\leftrightarrow\ket{00}, \quad &\ket{2}&\leftrightarrow\ket{01}, \\
&\ket{3}\leftrightarrow\ket{10}, \quad &\ket{4}&\leftrightarrow\ket{11}, \\
&\ket{5}\leftrightarrow\ket{20}, \quad &\ket{6}&\leftrightarrow\ket{21}, \\
&\ket{7}\leftrightarrow\ket{\mathrm{ancillary}}.\\
\end{aligned}
\end{equation}
Here $\ket{n}$ stands for $n$th energy level, and $\ket{mk}$ for the basis states of the qutrit-qubit system $\ket{m}_A\otimes\ket{k}_B$.
We note that the ancillary level is assumed to be not occupied during operations and is employed for realization of arbitrary operators from $\mathrm{U(6)}$ group via operators from $\mathrm{SU(7)}$ group~\cite{Kiktenko2}.

From a physical point of view, quantum gates can be realized by a sequence of laser pulses.
In the resonance case, where the frequency of an electromagnetic pulse coincides with the frequency of transition between $\ket{n}$ and $\ket{m}$ levels ($m>n$), 
we obtain the following Hamiltonian in the interaction picture: 
\begin{equation}
	\mathcal{V}_{mn}(E)=Ed_{nm}\ket{n}\bra{m}+E^{*}d^*_{nm}\;\ket{m}\bra{n}, 
\end{equation}
where $E$ is a complex field amplitude and $d_{nm}$ is a transition dipole moment:
\begin{equation}
	d_{nm}=\bra{\psi_n}(qx)\ket{\psi_m}=\bra{\psi_n}x\ket{\psi_m}.
\end{equation}
Exact calculations show that dipole moments are equal to zero for transitions without parity switching, 
and for transitions between adjacent energy levels they are much higher then others. 
Therefore, one can consider $m=n+1$ term only.

The additional term in the Hamiltonian generates the time evolution of the system, described by the operator
\begin{equation}
	U_{nm}(t,E)=\exp\left[-i \mathcal{V}_{nm}(E) t\right].
\end{equation}

By opting parameters of the field (amplitude, phase, and pulse duration), we can obtain the evolution operator a required form. 
For convenience of further description, we introduce operator that performs rotation around $Y$-axis of the Bloch sphere of the particular two-dimensional Hilbert subspace:
\begin{equation}
\begin{split}
	R_{nm}(\theta)&=U_{nm}\left(\frac{\theta}{2E_0d_{nm}},E_0\exp\left[i\frac{3\pi}{2}\right]\right)=\\
	&=\left[ \begin {array}{cc} 
	\cos(\theta/2)&-\sin(\theta/2)\\
	\sin(\theta/2)&\cos(\theta/2)
	\end {array} \right]^{(nm)}\bigoplus\mathbb{I}_5^{(\overline{nm})},
\end{split}
\end{equation}
where $E_{0}$ is an arbitrary real field amplitude, 
the matrix subscript $(n,m)$ indicates that it is written in the basis $\{\ket{n},\ket{m}\}$, $\bigoplus$ stands for the direct sum, 
$\mathbb{I}_5$ stands for the identity operator in $5$-dimensional Hilbert space and superscript $(\overline{nm})$ indicates that the identity operator acts in the orthogonal complement 
$(\mathrm{Span}\{\ket{n},\ket{m}\})^{\perp}$.

A possible scenario of using such a system for quantum computing can be demonstrated in the parity check problem~\cite{Parity1,Parity2,Parity3}. 
The problem can be formulated as follows.
We consider a sequence of six Boolean variables 
\begin{equation}
	\mathbf{s}=\{s_1,s_2,\ldots,s_6\}, \quad s_i\in\{0,1\},
\end{equation} 
and we are asking whether the string $\mathbf{s}$ is even or odd. 
Strictly speaking, this question can be formulated using the parity function
\begin{equation}
	p(\mathbf{s})=s_1 \oplus s_2 \oplus s_3 \oplus s_4 \oplus s_5 \oplus s_6,
\end{equation}
where $\oplus$ denotes the mod 2 summation.

To cope with this task a classical computer needs to determine the value of each bit from the string, \emph{i.e.} to have six references to the string.
The quantum computer can solve this problem with only three references to the string by using special operator (oracle):
\begin{equation}
\mathbf{O_s}\ket{nm}=(-1)^{s_{2n+m+1}}\ket{nm},
\end{equation}
where $n=0,1,2$ and $m=0,1$.
Here we note that the current algorithm with the sting of length two is equivalent to the Deutsch algorithm~\cite{Deutsch}.

An algorithm that solves parity problem using qubit--qutrit system is shown in Fig.~\ref{fig:algorithm}a. 
It can be represented as an operator of the following form:
\begin{equation}
\mathbf{G}=\mathbf{H}^{(B)}\mathbf{O_s}\mathbf{U}^{(A)}_{12}\mathbf{O_s}\mathbf{U}^{(A)}_{01}\mathbf{O_s}\mathbf{H}^{(B)}.
\end{equation}
Here $\mathbf{H}^{(B)}$ is a Hadamard gate, acting on a qubit; $\mathbf{U}^{(A)}_{nm}$ is a gate, 
that performs swap $\ket{m}\leftrightarrow\ket{n}$ in a qutrit. 
The initial state of the system is $\ket{00}$, and it can be shown, that $\mathbf{G}$ acts on it as
\begin{equation}
\mathbf{G}\ket{00}=\alpha\left[(1-p(\mathbf{s}))
\ket{20}+
p(\mathbf{s})\ket{21}\right],
\end{equation}
where global phase $\alpha=(-1)^{s_1+s_3+s_5}$.
All required gates can be realized as sequences of laser pulses (see Fig. \ref{fig:algorithm}b) as follows:
\begin{equation}
\begin{split}
	\mathbf{H}^{(A)}=&R_{67}(2\pi)R_{56}\left(\frac{7\pi}{2}\right) R_{34}\left(\frac{5\pi}{2}\right)\\
	&R_{23}(2\pi)R_{12}\left(\frac{7\pi}{2}\right),\\
	\mathbf{U}^{(B)}_{01}&=R_{23}(\pi)R_{34}(\pi)R_{12}(\pi)R_{23}(\pi),\\ 
	\mathbf{U}^{(B)}_{12}&=R_{45}(\pi)R_{56}(\pi)R_{34}(\pi)R_{45}(\pi),\\ 
	\mathbf{Z}_n&=\prod\nolimits_{m=n}^{6}R_{m\;m+1}(2\pi),\\ 
	\mathbf{O_s}&=\prod\nolimits_{n=1}^{6}\mathbf{Z}_n^{s_n}
\end{split} 
\end{equation}
(the full form of operators in the space of seven-level system is presented in the Appendix A).

The result of an algorithm is read by the measurement of the qubit $B$ in the computational basis. 
It can be described by the following observable acting in the space of the system: 
\begin{equation}
\begin{split}
	\sigma_z^{(B)}=&(\ket{1}\bra{1}+\ket{3}\bra{3}+\ket{5}\bra{5})\\
	-&(\ket{2}\bra{2}+\ket{4}\bra{4}+\ket{6}\bra{6}).
\end{split}
\end{equation}
At the same time it turns out to be the measurement of the parity of the wavefunction of the particle: 
even functions correspond to even strings and odd functions correspond to odd strings. 
From the practical point of view, this measurement also can be implemented by measuring probability density in the middle of the square-well trap (that is zero for odd functions).

\section{Conclusion}\label{sec:conclusion}

To conclude, we point out the main results of our work.
We studied a seven-level quantum system as a platform for realization of quantum information processing.
We employed the three-qubit encoding (\ref{eq:rerp1}) in order to demonstrate the subadditivity and strong subadditivity conditions in the thermal state of the particular noncomposite system.
In the result nonmonotonic behaviour of correlations as function of the temperature in particular partition was revealed.
The qubit--qutrit encoding (\ref{eq:repr2}) was used in order to suggest a single qudit algorithm for calculation of the parity of a 6-bit string.
As a possible experimental setup, a particle in a 1D square-well trap,
which can be realized using currently available experimental tools, was considered.

\section*{Acknowledgments}
The support from Ministry of Education and Science of the Russian Federation in the framework of the Federal Program (Agreement 14.579.21.0105, ID RFMEFI57915X0105) is acknowledged. 

\begin{widetext}
\section*{Appendix A}
Here we present the explicit form qubit and qutrit gates from the viewpoint of seven-dimensional Hilbert space of noncomposite system:
\begin{equation}
	\mathbf{O_s}=\begin{bmatrix}
		(-1)^{s_1}&0&0&0&0&0&0\\ 
		0&(-1)^{s_2}&0&0&0&0&0\\ 
		0&0&(-1)^{s_3}&0&0&0&0\\ 
		0&0&0&(-1)^{s_4}&0&0&0\\ 
		0&0&0&0&(-1)^{s_5}&0&0\\ 
		0&0&0&0&0&(-1)^{s_6}&0 \\
		0&0&0&0&0&0&(-1)^{\sum_{m=1}^{6}s_m} \\
	\end{bmatrix},
\end{equation}
\begin{equation}
	\mathbf{H}^{(A)}=
	\begin{bmatrix}
	\frac{\sqrt{2}}{2}&\frac{\sqrt{2}}{2}&0&0&0&0&0\\
	\noalign{\medskip}\frac{\sqrt{2}}{2}&-\frac{\sqrt{2}}{2}&0&0&0&0&0\\
	\noalign{\medskip}0&0&\frac{\sqrt{2}}{2}&\frac{\sqrt{2}}{2}&0&0&0\\
	\noalign{\medskip}0&0&\frac{\sqrt{2}}{2}&-\frac{\sqrt{2}}{2}&0&0&0\\
	\noalign{\medskip}0&0&0&0&\frac{\sqrt{2}}{2}&\frac{\sqrt{2}}{2}&0\\
	\noalign{\medskip}0&0&0&0&\frac{\sqrt{2}}{2}&-\frac{\sqrt{2}}{2}&0\\
	\noalign{\medskip}0&0&0&0&0&0&-1
\end{bmatrix},
\end{equation}
\begin{equation}
	\mathbf{U}^{(B)}_{01}=\begin{bmatrix}
	0&0&1&0&0&0&0\\ 
	0&0&0&1&0&0&0\\ 
	1&0&0&0&0&0&0\\ 
	0&1&0&0&0&0&0\\ 
	0&0&0&0&1&0&0\\ 
	0&0&0&0&0&1&0\\
	0&0&0&0&0&0&1
\end{bmatrix},
\quad
\mathbf{U}^{(B)}_{12}=
\begin{bmatrix}
	1&0&0&0&0&0&0\\0&1&0&0&0&0&0\\0&0&0&0&1&0&0\\0&0&0&0&0&1&0\\0&0&1&0&0&0&0\\0&0&0&1&0&0&0\\0&0&0&0&0&0&1
\end{bmatrix}.
\end{equation}
One can see that the seventh level is used to accumulate the phase, keeping the whole operation in SU(7) group.

\end{widetext}

\end{document}